\documentclass{emulateapj}

\newcommand{\pubjournal}[6] {#1 #5, #2, {\bf #3}, #4}

\shorttitle{DM burning effects on PopIII}
\shortauthors{Fabio Iocco}

\begin{document}

\title{Dark Matter capture and annihilation on the First Stars: preliminary estimates}
\author{Fabio Iocco}
\affil{
INAF/Osservatorio Astrofisico di Arcetri, Largo Enrico Fermi 5, Firenze, 50125, Italy\\
   Kavli Institute for Particle Astrophysics and Cosmology, Stanford University, Stanford, CA 94309}
\email{iocco@arcetri.astro.it}

\begin{abstract}
Assuming that Dark Matter is dominated by WIMPs, it accretes by gravitational attraction and scattering over baryonic material and annihilates inside celestial objects, giving rise to a ``Dark Luminosity'' which may potentially affect the evolution of stars.
We estimate the Dark Luminosity achieved by different kinds of stars in a halo with DM properties characteristic of the ones where the first star formation episode occurs.
We find that either massive, metal-free and small, galactic-like stars can achieve Dark Luminosities comparable or exceeding their nuclear ones.
This might have dramatic effects over the evolution of the very first stars, known as Population III.
\end{abstract}


\section{Introduction}
Observations continue to provide evidence for a {\it dark} component of matter in the Universe, whose motivations and possible candidates are thouroughly reviewed by \cite{Bertone:2004pz}.
It has been realized quite early that scattering between Dark Matter (DM) and baryons would result in capture of DM particles over celestial bodies, with the most remarkable effect for stars to host within their bosom an additional source of energy due to the captured DM annihilation. 
Since then, several authors have studied the possibility to exploit the effects of an additional, potentially unexhaustable, energy source within stars in order to constraint DM properties: among the latest, \cite{Moskalenko:2007ak} and \cite{Bertone:2007ae} study compact objects; \cite{Fairbairn:2007bn} and \cite{Scott:2007md} have performed preliminary numerical calculations of the stellar evolution. 
All the results seem to converge over the fact that within a neutralino dominated DM scenario, and with an astrophysical environment with typical galactic DM velocity profiles of the order of $\bar v\approx$10$^2$ km/s, DM densities required in order to have appreciable effects over stellar evolution are $\rho_\chi\geq$10$^7$GeV cm$^{-3}$, achievable only within the central two parsecs of our Galaxy, with similar restrictions holding for other types of galaxies: DM burning would not affect most of the stars in our local Universe.
However, so far no one has studied the effects of DM burning over the first stars in the Universe: according to the current $\Lambda$CDM cosmological description the early Universe is quite a peculiar place for stellar formation to take place. The almost total absence of metals, \cite{Iocco:2007km}, the small mass and angular momentum of the young and dense haloes where the first stars form, the lack of strong magnetic fields, all intertwine together giving the Population III peculiar properties; above all, is to be mentioned the characteristic formation of a single, massive (30M$_\odot$-300M$_\odot$), \cite{Abel:2001pr} and \cite{Gao:2006ug}.
Massive, metal free stars would evolve quickly through the Main Sequence (MS), burning all of their fuel within timescales of 10$^6$ years and proceeding quickly through the following element burning, until at the oxygen burning stage temperature and entropy are so high that copious e$^+$e$^-$ production takes place, pressure drops and a runaway collapse is triggered. This leads to explosive nucleosynthesis which results in direct collapse to a Black Hole, or to a direct dramatic explosion, or to stellar pulsations followed by a final violent explosion, \cite{Heger:2001cd}.
An ``alteration'' of the basic stellar formation scenario which has been pointed out by \cite{Spolyar:2007qv}: at times much earlier than the actual formation of a protostar (T$\leq$10$^4$K, $n\leq$10$^{18}$ cm$^{-3}$) Dark Matter embedded in the halo annihilates at a rate such that the energy deposited exceeds the one loss by the molecular cooling of the gas. Although these semianalytical results have to be confirmed by more detailed numerical simulations it is clear that this process could change the history of primordial stellar formation by preventing the cloud from collapsing, or by slowing it down or else inducing fragmentation of the baryonic cloud. Although this process is different from the one we study it is clearly related to it: when the cloud becomes opaque enough to efficiently scatter WIMPs off baryons another energy source would be added in the center, as opposed to the pure DM annihilation, which diffuse everywhere in the halo. 
Here we present preliminary estimates of the DM accretion and annihilation inside a star, as it would result is it had formed in ``classical'' way, without any effect from the DM annihilation at lower densities and temperatures.
\section{Dark Matter capture and annihilation in stars}\label{equations}
Models of DM capture and annihilation on celestial bodies were developed in the 80s by many authors, and here we collect their results of most concern for this Letter.
At the equilibrium between capture and annihilation the event rate is, after \cite{gould87}:
\begin{equation}
C=4\pi \int_0^{R_*} dr\, r^2\, \frac{dC(r)}{dV},
\label{gould2.27}
\end{equation}
where

\begin{eqnarray}
\frac{dC(r)}{dV}&=&
\left(
\frac{6}{\pi}
\right)^{1/2}
\sigma_0 A_n^4 \frac{\rho_*}{M_n}\frac{\rho_\chi}{m_\chi}
\frac{v^2(r)}{\bar{v}^2} \frac{\bar{v}}{2\eta A^2} 
\label{gould2.24}\\
&\times&
\left\{
\left(
A_+A_- -\frac12
\right)
\left[
\chi(-\eta,\eta) -\chi(A_-,A_+)
\right]
\right.\nonumber\\
&+&\left.
\frac12A_+e^{-A_-^2} -\frac12A_- e^{-A_+^2}-\eta e^{-\eta^2}
\right\},\nonumber\\
A^2&=& \frac{3v^2(r)\mu}{2\bar{v}^2\mu_-^2},   \  \  A_\pm=A\pm\eta,  \  \   \eta=\sqrt \frac{3v_*^2}{2\bar{v}^2}, \nonumber\\
\chi(a,b)&=&\int_a^b dy\, e^{-y^2}=
\frac{\sqrt\pi}{2}[{\rm erf}(b)-{\rm erf}(a)],
\nonumber
\label{Caprate}
\end{eqnarray}
$\rho_\chi$ is the ambient WIMP energy density, $A_n$ is the atomic number of the star's nuclei, $M_n$ is the nucleus mass, $\bar{v}$ is the WIMP velocity dispersion, $v_*$ is the star's velocity with respect to an observer and $\mu=m_\chi/M_n$, $\mu_-=(\mu-1)/2$, $v(r)$ is the escape velocity at a given radius $r$ inside a star, and subscript $*$ refers to stellar quantities.
The capture-annihilation process will reach the equilibrium on timescales $\tau_\chi$ with most of the particles concentrated within a radius $r_\chi$, which read respectively:
\begin{equation}
\tau_{\chi}=\left(\frac{C\langle\sigma v\rangle}{\pi^{3/2}r^3_\chi}\right)^{-1/2}, \ \ r_{\chi}=c\left(\frac{3kT_c}{2\pi G\rho_c m_\chi}\right)^{1/2};
\label{Eqtimescale} 
\end{equation}
$T_c$ and $\rho_c$ are the stellar core temperature and density, respectively, as shown in \cite{gs87}. According to them, \emph{inside} the star the captured DM follows a density profile $n_\chi$, with a central maximum density $n_\chi^c$, respectively described by:
\begin{equation}
n_{\chi}(r)=n^c_\chi\exp(-r^2/r_\chi^2), \ \  n_\chi^c=\frac{C\tau_\chi}{\pi^{3/2}r^3_\chi}.
\label{DMprofile} 
\end{equation}
Also, DM particles can transport energy by scattering off nuclei inside the star, thus constituting an alternative transport method inside the star; after \cite{Spergel:85}, the quantity of energy transported by the WIMPs per unit volume per unit time, $\epsilon_\chi$, reads:
\begin{eqnarray}
\epsilon_\chi&=&8\sqrt{\frac{2}{\pi}}n_\chi n_n \sigma_0 \frac{m_\chi m_n}{(m_\chi+m_n)^2} 
\nonumber\\
& &
 \left(\frac{m_n k T_\chi + m_\chi k T_n}{m_\chi m_n}\right)^{1/2}k(T_n-T_\chi),
\label{Entransp}
\end{eqnarray}
where the subscript $n$ refers to values for the nucleons.
$L_\chi$=$m_\chi C$ depends neither on $\langle\sigma v\rangle$ nor on the neutralino mass $m_\chi$, whereas they both affect the distribution of the DM inside the star and the transport effects. 
It is worth noticing that although the product of an annihilation (of the dark matter accreted inside the star), $L_\chi$ depends on \emph{environmental} quantities of dark matter (environment density, velocity dispersion) with the typical properties of a scattering. Also, both $L_\chi$ and $\tau_\chi$ are degenerate in the product ${\cal D}$=$\sigma_0 \times \rho_\chi$/$\bar v$ (actually, plus an additional dependence on $\bar v$, which is a posteriori negligible for this illustrative purpose), which we will use as a parameter in the following as it conveniently summarizes the DM values in the halo, with $\sigma_0$.
\section{Application to early stars: preliminary results}\label{applstars}
The DM density and its velocity dispersion play a fundamental role in determining the accretion rate over any stellar object; in the local Universe, high DM densities are achieved only in the very centre of galaxies; in particular in the Milky Way, a cuspy profile is believed to have formed by accretion around a central Black Hole, over the long lifetime of our galaxy and its central object, see for instance in \cite{bm05} and references therein. DM particles adopt Keplerian velocities around the central object, thus reaching their highest velocities where the DM density peaks; on the other hand, most of the stellar mass is not located in the center of the Galaxy, but in regions where the DM density is too small to consistently affect stellar evolution. 

In a primordial environment the scenario is quite different: in a young halo such as the ones where PopIII stars form average densities are quite high, as the concentration of Dark Matter, $c$, is higher in a younger Universe. Moreover, no central object exists yet so DM particles do not have time to develop a keplerian velocity profile: the star itself is believed to be the very first object to form in the center of the halo (where the DM density is highest). In a primordial halo, for the whole lifetime of a PopIII star, we can assume DM dispersion velocities to be a Maxwell-Boltzmann distribution around the central value corresponding to the virial temperature of the halo itself.
In order to estrapolate the values of the DM density $\rho_\chi$ at the location of a primordial star, we quote state of the art simulations of the collapse of the the first object. They achieve impressive resolutions, having to be stopped at central gas (baryon) densities of ${\cal O}$(10$^{25}$ particles/cm$^3$), achieved at a radii of $\approx$10$^9$cm, \cite{TurkFS3}; unfortunately this is not (yet) enough to resolve the actual {\it star}; simulations stop following DM particles even earlier, when the density is homogenous within a radius of $\approx$10$^{14-15}$cm, thus not allowing us to have more detailed insights on smaller scales. 
For a central region of radius 10$^{15}$cm one finds an enclosed mass M$_{DM}\approx$10M$_\odot$, corresponding to a $\rho_\chi\approx$10$^{12}$GeV cm$^{-3}$; this data, quoted from the state of the art simulations \cite{TurkFS3} and \cite{Turkpriv}, is in agreement with the results extrapolated from previous simulations and also with the ones obtained by much simpler semianalitical models of adiabatically contracted DM halos, as in \cite{Spolyar:2007qv}. 
Namely, the halo followed in Turk's cosmological simulations has M$_{Halo}$=${\cal O}$(10$^6$M$_\odot$) a virial temperature T$\approx$10$^3$K and therefore a virial velocity $\bar v\approx$10$^6$cm s$^{-1}$, according to mass-temperature relations evolution with redshift in \cite{Levine:2002uq} and references therein.
It is likely that the final $\rho_\chi$ in the center of the halo will be higher at the time of stellar formation; moreover, the effects triggered by a process such as the one discussed in \cite{Spolyar:2007qv} are not trivial to predict. Here we limit ourselves to present results using $\rho_\chi$=10$^{12}$GeV cm$^{-3}$ as our fiducial value, adopting $\langle \sigma v \rangle$=3$\times$10$^{-26}$cm$^3$s$^{-1}$ after \cite{Bertone:2004pz} and $\sigma_0$=10$^{-38}$(/10$^{-43}$)cm$^2$ for the spin-dependent(/independent) case, \cite{cdms06}/\cite{sk04}; this corresponds to a fiducial value for our parameter ${\cal D}$=10$^{-32}$(/10$^{-37}$)GeV s/cm$^2$ for the spin dependent(/independent) case. Results for different values can be easily rescaled.

\begin{table}
\begin{center}
\begin{tabular}{|l|l|l|l|l|l|}
\hline A$_n$& $L_\chi$(erg/s) & $\tau_\chi$(s) & r$_\chi$(cm) & $n_\chi^c$(GeV/cm$^3$) & $\epsilon_\chi$(erg/s/cm$^3$)\\
  \hline 1 &10$^{40}$ & 10$^{7}$ & 10$^{10}$ & 10$^{18}$& 10$^3$\\
  \hline 4 &10$^{38}$ & 10$^{8}$ & 10$^{9}$ & 10$^{17}$& -10$^2$\\
\hline
 \hline  1 &10$^{38}$ & 10$^{10}$ & 10$^{11}$ & 10$^{16}$ & 1\\
\hline 
\end{tabular}
\caption{Values for a 75 M$_\odot$, initial metallicity Z=10$^{-4}$, in a neutralino case with $m_\chi$=100GeV. ${\cal D}$=10$^{-32}$(/10$^{-37}$)GeV s/cm$^2$ for the spin dependent(/independent) case. Last line refers to the H shell during helium burning.}\label{Tab1}
\end{center}
\end{table}

\begin{table}
\begin{center}
\begin{tabular}{|l|l|l|l|l|l|}
\hline M$_*$(M$_\odot$) & $L_\chi$(erg/s) & $\tau_\chi$(s) & r$_\chi$(cm) & $n_\chi^c$(GeV/cm$^3$) & $\epsilon_\chi$(erg/s/cm$^3$)\\
\hline 1 &10$^{38}$ & 10$^{7}$ & 10$^{9}$ & 10$^{18}$& 10$^4$\\
\hline 13 &10$^{39}$ & 10$^{7}$ & 10$^{10}$ & 10$^{18}$& 10$^2$\\
\hline 25 &10$^{40}$ & 10$^{7}$ & 10$^{10}$ & 10$^{18}$&10$^2$\\
\hline
\end{tabular}
\caption{Same as in Table \ref{Tab1}, for MS low-mass stars.}\label{Tab2}
\end{center}
\end{table}
Using the equations introduced in Section \ref{equations}, we have estimated those quantities for WIMPs captured by a 75M$_\odot$, extremely metal-poor (Z=10$^{-4}$Z$_\odot$) star, described in \cite{woosley02}; our results do not change within the order of magnitude if we adopt the values of a metal-free, 100M$_\odot$ star described in \cite{Marigo:2001pm}.
In fact, stellar radii of ${\cal O}$ (R$_* \approx$10R$_\odot$) during the MS and R$_*\approx$10$^{3}$R$_\odot$cm during the helium burning phases are achieved both by the 75M$_\odot$ star in \cite{woosley02} and the metal-free 100M$_\odot$ one in \cite{Marigo:2001pm}, as it can be seen from a detailed table in the first case and extrapolated from the diagram on page 5 in the second.
While this paper was being edited for publication, I have received values of Zero Age Main Sequence, metal-free stars from P.~Marigo (unpublished results after \cite{Marigo:2001pm}): Table \ref{Tab3} contains the relevant quantities calculated as from equations in Section \ref{equations} for these actual metal-free stars;  $\tau_\chi\approx$10$^6$s, $n_\chi^c$=10$^{19}$GeV/cm$^3$, and $\epsilon_\chi\approx$10$^6$ erg/s/cm$^3$ for the whole mass range; $L^{ZAMS}_*$ is the nuclear luminosity of the star, $R_*$ is its radius in cm, which we present for direct comparison with $r_\chi$. 
For the case of the 75M$_\odot$ star, in order to estimate the mass of the core for the stages following the H burning, and therefore the central mass of the star which has a definitely different composition, we use the approximate relation from \cite{Heger:2001cd} (which has to be taken with care in this case, as we are out of the PISNe range) M$_{He}$=13/24(M$_{ZAMS}$-20), thus obtaining an helium core for the 75M$_{\odot}$ star of M$_{He}\approx$30M$_{\odot}$. During the helium burning stage the star's central density is $\rho^{He}_c\approx$3$\times$10$^2$g/cm$^3$.
Approximating the core with a constant density of $\rho^{He}$=10$^2$g/cm$^3$ one gets a central He core of radius R$^{He}\approx$5$\times$10$^{10}$cm (${\cal O}$(R$_\odot$)). The remaining 45M$_\odot$ of hydrogen will continue accreting DM (with a spin-independent cross section, while the helium core accretes it with a spin-dependent one). 
In Table \ref{Tab1} we report the quantities inroduced in Section \ref{equations} as for the 75M$_\odot$ hydrogen star, as described so far, for a neutralino with mass $m_\chi$=100GeV. 
\begin{table}
\begin{center}
\begin{tabular}{|l|l|l|l|l|}
\hline M$_*$(M$_\odot$) & $L_\chi$(erg/s) & r$_\chi$(cm) & $R_*$(cm)&$L_\chi$/$L^{ZAMS}_*$ \\
\hline 50 &4$\times$10$^{40}$ &2$\times$10$^{9}$ &2$\times$10$^{11}$& 25\\
\hline 70 &7$\times$10$^{40}$ &3$\times$10$^{9}$ &2$\times$10$^{11}$& 22\\
\hline 100 &1$\times$10$^{41}$ &4$\times$10$^{9}$&3$\times$10$^{11}$&21\\
\hline 200 &3$\times$10$^{41}$ &5$\times$10$^{9}$ &4$\times$10$^{11}$& 20\\
\hline 300&5$\times$10$^{41}$ &5$\times$10$^{9}$ &5$\times$10$^{11}$& 21\\
\hline 500 &1$\times$10$^{42}$ &6$\times$10$^{9}$&7$\times$10$^{11}$& 23\\
\hline 600 &2$\times$10$^{42}$ &6$\times$10$^{9}$&8$\times$10$^{11}$&24 \\
\hline
\end{tabular}
\caption{Same as in Table \ref{Tab1}, for ZAMS metal--free stars.}\label{Tab3}
\end{center}
\end{table}
The luminosity of a 75M$_\odot$ star is of the order of 10$^6$L$_\odot\approx$10$^{39}$erg s$^{-1}$ throughout all its life, until the oxygen burning stage, as for instance in \cite{Marigo:2001pm}.
Our estimates show that a Dark Luminosity 10$^{37}$erg s$^{-1}\leq$ $L_\chi$ $\leq$10$^{43}$erg s$^{-1}$ is achieved for DM densities in the range [10$^9$GeV cm$^{-3}$, 10$^{15}$GeV cm$^{-3}$].
These values are extremely interesting, as they are comparable (actually exceed, at the cross section upper limit) the stellar luminosity during the Main Sequence.

Transport effects:
an upper limit can be obtained by using Eq. \ref{Entransp}; the highest temperatures WIMPs can achieve correspondence to the escape velocity where most of the WIMPs are concentrated, namely within the radius r$_\chi$.  
By setting the WIMP temperature with the escape velocity at r$_\chi$, and number densities n$_\chi$ and n$_p$ for WIMPs and protons respectively, one gets values fo $\epsilon_\chi$ as reported in the Tables.
The DM transport effects in the core seems to be negligible for the whole neutralino density and mass range studied: the stellar luminosity L$_*$=10$^{39}$ erg/s needs an efficiency $\epsilon_* \approx$10$^9$ erg s$^{-1}$ cm$^{-3}$, if one assumes a nuclear core of 10$^{10}$cm, two orders of magnitude more than upper limit value of $\epsilon_\chi$ achieved with the highest DM density considered in our range.
During the helium burning, $\tau_\chi$ is to be read in a very indicative way: the higher DM density accumulated during the MS may simply consume itself until it reaches a lower concentration at the equilibrium in each part of the star. However, we find it interesting that a capture rate equivalent to the helium core is achieved by the huge hydrogen envelope surrounding it.

In Table \ref{Tab2} we show results for low-mass stars; although the formation of low-mass stars within the Population III it is not the currently favored hypothesis, state of the art simulations cannot totally rule out a low-mass peak in the IMF. Some authors, as for instance \cite{Choudhury:2006bp}, do actually invoke a low-mass, Salpeter-like, PopulationIII. 

These values are very interesting, too: as already mentioned, other authors have already studied the evolutionary behavior of low-mass stars in presence of a DM annihilation source, and more detailed calculations are expected to be available soon. We only stress that for low-mass stars the behavior of metal-free objects is not so dramatically different as for massive stars, and that results obtained for metal-rich, ``usual'', stars can be used at least for a first approximation understanding of the behavior of low-mass Population III stars in presence of a relevant Dark Luminosity.

\section{Discussion}\label{discussion}
Without entering detailed calculations, we wish to focus the attention on several issues that could be raised by an ``unorthodox'' behavior of massive PopIII stars, hoping this Letter will be used as a starting point for more detailed and extensive analysis.
The formation of primordial stars has to be carefully investigated in the presence of Dark Matter annihilation, according to the preliminary, interesting results obtained by \cite{Spolyar:2007qv}; if slowed down, the cloud might start capturing DM at an early stage, and Dark Luminosity might start playing a role even before the formation of the actual star, if the cloud collapse is not stopped by this process.
Assuming that a primordial star forms in a ``standard'' fashion, and lately develops Dark Luminosities of the order of our upper limits, there is no telling, without actual simulations and detailed calculations, what the behavior of the object could be (we recall that for the 75M$_\odot$ our fiducial value for ${\cal D}$=10$^{-32}$GeV s/cm$^2$ is $L_\chi$=10$^{40}$erg/s, one order of magnitude greater than the stellar luminosity).
In the likely case that the actual value of $\sigma_0$ is not its quoted upper limit, still some interesting scenarios can be envisioned:
PopIII stars are thought to explode as PISNe, due to the high entropy and temperature of the core at the time of oxygen burning; what if a Dark Luminosity comparable to the stellar one can reduce both entropy and temperature of the star, thus leading it through the oxygen burning phases without dramatic instabilities? What would the effects of an $L_\chi$ only comparable with the required stellar luminosity over the short {\it pp} phase at the beginning of the hydrogen burning forced by the lack of CNO elements, given the very slow {\it pp} energy production sensitivity from temperature?
At the same time, stellar evolution in presence of an additional energy source might lead to very different nucleosynthetic yields; the odd-Z element deficiency has been long considered one of the ``signatures'' of PopIII stars, what if different temperatures (plus an additional electron source due to neutralino decay) would result in a different peculiar signature?
What would happen to a star which is supported mainly by DM burning, and would survive long enough to follow its host halo in a merger, until DM environmental conditions change? 
PopIII stars are thought to be the first engine for the Reionization of the Universe, as they are a huge source of ionizing photons as a consequence of their very high surface temperature; can Dark Luminosity modify it, and thus result in a modification of the Reionization models?

It is straightforward to ask whether a clear signature of this process could be recognized. High energy neutrinos produced by DM annihilation inside the Sun are expected to be observed in the future by IceCube, \cite{Mena:2007ty}; the same process would lead to a diffuse high energy neutrino background from the first stars. However, a massive star burning DM for $\tau_*\approx$10$^6$ years with $L_\chi$=10$^{40}$ erg/s, would release a total energy $E_\chi\approx$10$^{54}$erg; let us take as an upper limit that all of this energy ends up in neutrinos (the only annihilation channel being $\chi\chi\rightarrow\nu\nu$), adopt the formalism described in \cite{Iocco:2007td} and the same stellar formation rate (the ``fiducial'' one as in \cite{Choudhury:2004vs}), and assume the formation of a single stellar object per halo. One gets a flux at the Earth, at the peak energy E$_\nu$=$m_\chi$/(1+$\bar z$)$\leq$10GeV ($m_\chi$=100GeV) of $\Phi\leq$10$^{-5}$ GeV cm$^{-2}$ sr$^{-1}$ s$^{-1}$ ($\bar z \geq$10 being the ``central'' redshift of PopIII episode); this flux is buried several orders of magnitude below the diffuse atmospheric neutrino background at the same energy as from \cite{Evoli:2007iy}, thus making neutrinos a non efficient tool for studying this process.
\section{Preliminary conclusions}\label{conclusion}
Within a WIMP dominated DM scenario, with structure formation taking place in a ``standard'' scenario, primordial stars accrete DM much more efficiently than most of modern, galactic stars, mainly due to the peculiar conditions of the halos where they form. Within a large region of the relevant parameter space (DM velocity dispersion, density and WIMP-baryon scattering cross-section) the energy deposited inside the star by the accreted DM annihilation is comparable to (and even exceeds) the stellar, nuclear luminosity. This raises several questions about the real nature of primordial stars, and about whether their behavior as stellar objects is dictated by baryons only.

Whereas describing the effects of DM ``burning'' on the first stars would, at this stage, be pure speculation, we aim with this Letter to stimulate discussion and interest on this topic, and qualitatively propose scenarios whose validity will have to be checked in the future. However, we think the questions raised by our preliminary estimates are extremely relevant for our understanding of first object formation and evolution in the Universe, which is becoming a much more complicated puzzle than previously expected.
\acknowledgments
The author is indebted to I.~V.~Moskalenko for introducing him to the topic of DM accretion in stars; he also kindly acknowledges fruitful conversations with S.~Akiyama, G.~Bertone and P.~D.~Serpico, and thanks G.~Miele for fundamental encouragements. This work is supported by the grant COFIN--MIUR Pacini 2006.

\end{document}